\input harvmac.tex
\let\includefigures=\iftrue
\newfam\black
\includefigures
\input epsf
\def\figin{\epsfcheck\figin}\def\figins{\epsfcheck\figins}
\def\epsfcheck{\ifx\epsfbox\UnDeFiNeD
\message{(NO epsf.tex, FIGURES WILL BE IGNORED)}
\gdef\figin##1{\vskip2in}\gdef\figins##1{\hskip.5in}% blank space instead
\else\message{(FIGURES WILL BE INCLUDED)}%
\gdef\figin##1{##1}\gdef\figins##1{##1}\fi}
\def\DefWarn#1{}
\def\figinsert{\goodbreak\midinsert}
\def\ifig#1#2#3{\DefWarn#1\xdef#1{fig.~\the\figno}
\writedef{#1\leftbracket fig.\noexpand~\the\figno}%
\figinsert\figin{\centerline{#3}}\medskip\centerline{\vbox{\baselineskip12pt
\advance\hsize by -1truein\noindent\footnotefont{\bf Fig.~\the\figno:} #2}}
\bigskip\endinsert\global\advance\figno by1}
%%%
\else
\def\ifig#1#2#3{\xdef#1{fig.~\the\figno}
\writedef{#1\leftbracket fig.\noexpand~\the\figno}%
%\figinsert\figin{\centerline{#3}}\medskip\centerline{\vbox{\baselineskip12pt
%\advance\hsize by -1truein\noindent\footnotefont{\bf Fig.~\the\figno:} #2}}
%\bigskip\endinsert
\global\advance\figno by1}
\fi
%\draftmode

\Title{\vbox{\baselineskip12pt\hbox{hep-th/0005129}
\hbox{CALT-68-2272}\hbox{CITUSC/00-023}}}
{\vbox{
\centerline{Space-Time Noncommutative}
\vskip 10pt
\centerline{Field Theories And Unitarity}
}}
\centerline{Jaume Gomis and Thomas Mehen}
%\medskip
\medskip
\medskip
\medskip
\centerline{\it Department of Physics}
\centerline{\it California Institute of Technology}
\centerline{\it Pasadena, CA 91125}
\centerline{\it and}
\centerline{\it Caltech-USC Center for Theoretical Physics} 
\centerline{\it University of Southern California}
\centerline{\it Los Angeles, CA 90089}

\medskip
\centerline{\tt gomis,\ mehen@theory.caltech.edu}
\medskip
\medskip
\medskip
%\bigskip
\noindent

We study the perturbative unitarity of noncommutative scalar field
theories. Field theories with space-time noncommutativity do not
have a unitary S-matrix. Field theories with only space
noncommutativity are perturbatively unitary. This can be understood from
string theory, since space noncommutative field theories describe a 
low energy limit of string theory in a background magnetic field. 
On the other hand, there is no regime in which space-time
noncommutative field theory is an appropriate description of string
theory. Whenever space-time noncommutative field theory 
 becomes relevant massive open string states
cannot be neglected.

\smallskip
%\medskip

\Date{May 2000}

\newsec{Introduction}

Noncommutative field theories are constructed 
from conventional (commutative) field
theories by replacing in the Lagrangian the usual multiplication of
fields with the $\star$-product of fields. The $\star$-product is
defined in terms of a real antisymmetric matrix $\theta^{\mu\nu}$ 
that parameterizes
the noncommutativity of Minkowski space-time\foot{Throughout the
paper we will use the $(+,-,\ldots ,-)$ convention for the signature 
of space-time.}
\eqn\comm{
[x^\mu,x^\nu]=i\theta^{\mu\nu}\qquad \mu,\nu=0,\ldots,D-1.}
The $\star$-product of two fields $\phi_1(x)$ and $\phi_2(x)$
is given by
\eqn\prod{
\phi_1(x) \star \phi_2(x) = e^{{i
\over 2} \theta^{\mu\nu} {\partial \over \partial \alpha^\mu}{\partial \over
\partial \beta^\nu}} \phi_1(x+\alpha) \,
\phi_2(x+\beta)|_{\alpha=\beta=0}.}
The noncommutativity in \comm\ gives rise to a space-time uncertainty
relation 
\eqn\uncert{
\Delta x^\mu\Delta x^\nu\ge {1\over 2}|\theta^{\mu\nu}|}
which leads to number of unusual phenomena such as the
mixing of the ultraviolet with the infrared as well as apparent
acausal behavior
\nref\mvs{S. Minwalla, M.V. Raamsdonk and N. Seiberg, ``Noncommutative
Perturbative Dynamics'', hep-th/9912072.}%
\nref\vs{M.V. Raamsdonk and N. Seiberg, `` Comments on Noncommutative
Perturbative Dynamics'', hep-th/0002186.}%
\nref\haya{M. Hayakawa,``Perturbative analysis on infrared
and ultraviolet aspects of
noncommutative QED on $R^4$,'' hep-th/9912167.}%
\nref\texa{W. Fischler, E. Gorbatov, A. Kashani-Poor, S. Paban,
P. Pouliot and J. Gomis, ``Evidence for winding states in
noncommutative quantum field theory,'' hep-th/0002067.}%
\nref\sunew{A. Matusis, L. Susskind and N. Toumbas,
``The IR/UV connection in the non-commutative gauge theories,''
hep-th/0002075.}%
\nref\ssta{N. Seiberg, L. Susskind and N. Toumbas, ``The Teleological
Behavior of Rigid Regge Rods'', hep-th/0005015.}%
\nref\chadempres{M. Chaichian, A. Demichev and P. Presnajder, ``Quantum
Field Theory on Noncommutative Space-Times and the Persistence of
Ultraviolet Divergences'',  Nucl. Phys. {\bf B567} (2000) 360,
hep-th/9812180; ``Quantum Field Theory on the Noncommutative Plane
with $E_q(2)$ Symmetry'', J. Math. Phys. {\bf 41} (2000) 1647, hep-th/9904132.}%
\mvs-\chadempres .

These field theories are non-local and this nonlocality has
important consequences for the dynamics
\mvs-\chadempres . The structure of the product
in $\prod$ leads to terms in the action with an infinite number of
derivatives of fields which casts some doubts on the 
unitarity  of noncommutative field theories. In this paper we will
check the unitarity of scalar noncommutative field theories at the
one loop level and show that theories with $\theta^{0i}=0$ are unitary
while theories with $\theta^{0i}\neq 0$ are not unitary. 

Noncommutative field theories with space noncommutativity
(that is
$\theta^{0i}=0$) have an elegant embedding in string theory
\nref\cds{A. Connes, M.R. Douglas and A. Schwarz, ``Noncommutative
Geometry and Matrix Theory: Compactification on Tori'', JHEP {\bf
9802}(1998) 003, hep-th/9711162.}%
\nref\dh{M.R. Douglas and C. Hull, "D-branes and the Noncommutative
Torus", JHEP {\bf 9802} (1998) 008, hep-th/9711165.}%
\nref\sw{N. Seiberg and E. Witten, `` String
Theory and Noncommutative Geometry'', JHEP {\bf 9909} (1999) 032,
hep-th/9908142.}%
\cds\dh\sw . They describe the low energy excitations of a D-brane 
 in the
presence of a background magnetic field\foot{We will
sometimes refer to these theories as magnetic theories.}. In this 
limit \sw\ the relevant
description of the dynamics is in terms of the noncommutative 
field theory of the massless open strings.
Both the massive open strings and the closed strings decouple\foot{In
\nref\ado{O. Andreev and H. Dorn, ``Diagrams of Noncommutative $\phi^3$
Theory from String Theory, hep-th/003113.}%
\nref\bcrso{A. Bilal, C.-S. Chu and R. Russo, ``String Theory and
Noncommutative Field Theories at One Loop'', hep-th/003180.}%
\nref\gkmrs{J. Gomis, M. Kleban, T. Mehen, M. Rangamani and S. Shenker,
``Noncommutative Gauge Dynamics From The String Worldsheet,
hep-th/0003215.}%
\nref\limi{H. Liu and J. Michelson, ``Stretched Strings in
Noncommutative Field Theory'', hep-th/0004013.}%
\nref\churosci{A. Bilal, C.-S. Chu, R. Russo and S. Sciuto,
``Multiloop String Amplitudes with B-Field and Noncommutative QFT'',
hep-th/0004183.}% 
\nref\kl{Y. Kiem and S. Lee, ``UV/IR Mixing in Noncommutative Field Theory
via Open String Loops'', hep-th/0003145.}%
\nref\rr{A. Rajaraman and M. Rozali, ``Noncommutative Gauge Theory, Divergences
and Closed Strings'', JHEP 0004:033 (2000).}%
\ado-\churosci\ one-loop string theory amplitudes were shown to
exhibit this decoupling. See also \kl-\rr .}. The consistent
truncation of the full unitary string theory to field theory with
space noncommutativity leads one to suspect that these field theories
are unitary. Moreover, these field theories are 
nonlocal in space but
are local in time. Therefore, a Hamiltonian can be constructed and it 
gives rise to  unitary time evolution of noncommutative  magnetic
field
 theories. 

Theories with space-time noncommutativity\foot{Likewise, we will
sometimes refer to these theories as electric theories.}
 (that is $\theta^{0i}\neq
0$) have an infinite number of time derivatives of fields in the
Lagrangian and  
are nonlocal in
time. The commutator in \comm\ leads to noncommutativity of the time
coordinate. 
Noncommutativity of the time coordinate and the corresponding
nonlocality in time results in theories where it is far from clear whether
the usual framework of quantum mechanics makes
sense. As such, noncommutative field theories with space-time
noncommutativity are excellent laboratories in which to test the
possible breakdown of the conventional notion of time or the familiar
framework of quantum mechanics in string
theory at the Planck scale\foot{In recent years, several examples of the 
breakdown of the conventional 
notion of space
at very short distances have been found in string theory such as  in
topology changing transitions
\nref\st{A. Strominger, ``Massless Black Holes and Conifolds in String
Theory'', Nucl. Phys. {\bf B451} (1995) 96, hep-th/9504090.}%
\nref\grs{B.R. Greene, D.R. Morrison and A. Strominger, ``Black Hole
Condensation and the Unification of String Vacua'', Nucl. Phys. {\bf B451}
(1995) 109, hep-th/9504145.}%
\st\grs .
It seems, therefore, imperative to
address the issue of possible breakdown of time.}. In this paper we will test
in these exotic field theories one of the basic
principles of quantum mechanics, the existence of a unitary
S-matrix. We explicitly show that several one loop amplitudes in
noncommutative scalar electric field theory are not unitary which demonstrates
that noncommutative field theories with space-time noncommutativity
clash with quantum mechanics. 

This field theory result meshes
very nicely with string theory expectations. 
$\theta^{0i}\neq 0$ is obtained by studying string theory in the
presence of a background electric field (recent work in this direction
has recently appeared in 
\nref\sstb{N. Seiberg, L. Susskind and N. Toumbas, ``Strings in
Background Electric Field, Space/Time \hskip-2pt Noncommutativity and
A \hskip-2pt New 
Noncritical String Theory'', \hskip-3pt hep-th/0005040.}%
\nref\gmms{R. Gopakumar, J. Maldacena, S. Minwalla and A. Strominger,
``S-Duality and Noncommutative Gauge Theory'', hep-th/0005048.}%
\nref\br{J.L.F Barb\'on and E. Rabinovici, ``Stringy Fuzziness as the
Custodial of Time-Space Noncommutativity'', hep-th/0005073.}%
\sstb-\br , see also
\nref\ggs{O.J. Ganor, G. Rajesh and S. Sethi, ``Duality and
Non-Commutative Gauge Theory'', hep-th/0005046.}%
\ggs).
There are three important parameters
that characterize the open strings
\nref\ft{E.S. Fradkin and A.A. Tseytlin, ``Nonlinear Electrodynamics
From Quantized Strings'', Phys. Lett. {bf 163B} (1985) 123.}%
\nref\clny{C.G. Callan, C. Lovelace, C.R. Nappi and S.A. Yost,
``String Loop Corrections To Beta Functions'', Nucl. Phys. {\bf B288}
(1987) 525; A. Abouelsaood, C.G. Callan, C.R. Nappi and S.A. Yost,
``Open Strings in Background Gauge Fields'', Nucl. Phys. {\bf B280}
(1987) 599.}%
\ft\clny\sw : $\alpha^\prime$, the metric
$G_{\mu\nu}$ and the noncommutativity matrix $\theta^{\mu\nu}$. One
must also keep in mind that there is an upper critical value on the
magnitude of the background electric field $E_c$
\nref\bur{C.P. Burgess, ``Open String Instability in Background
Electric Fields'', Nucl. Phys. {\bf B294} (1987) 427.}%
\nref\bapo{C. Bachas and M. Porrati, ``Pair Creation of Open Strings
in an Electric Field'', Phys. Lett. {\bf B296} (1992) 11.}%
\nref\nest{V.V. Nesterenko, ``The Dynamics of Open Strings in a
background Electromagnetic Field'', Int. J. Mod. Phys. {\bf A4} (1989)
2627.}%
\bur-\nest , beyond this value  the string
vacuum becomes unstable. It can be shown \sstb\gmms\br\  
using the relation between
these open string parameters with  the closed string
metric and background electric field that it is impossible to take a
consistent 
limit of string theory in
which $\theta^{\mu\nu}$ and $G_{\mu\nu}$ are kept fixed 
while $\alpha^\prime\rightarrow 0$. Therefore, unlike the case of
strings in a background magnetic field, it is impossible to find a
limit of string theory 
in which one is  left only with
a noncommutative field theory with fixed background metric
$G_{\mu\nu}$ and space-time noncommutativity parameter
$\theta^{0i}$. It is possible to find a limit of string theory 
\sstb\gmms\ with nonvanishing $\theta^{0i}$ in which the closed
strings 
decouple. However, 
$\theta^{0i}\sim \alpha^\prime$ making it impossible to decouple
massive open string states and keep $\theta^{0i}$ finite.
Thus, there is no sense in which the electric field
theories give an approximate description of a limit of string
theory. The lack of decoupling of the massless open string modes from
the massive ones gives very strong indication that the
noncommutative field theory
truncation  to the massless modes is  not unitary. This is 
consistent with what we find from
our field theory analysis.

The paper is organized as follows. In section $2$ we compute several
one loop amplitudes in noncommutative scalar field theory and show
that Feynman diagrams of space-time noncommutative theories do not
satisfy the usual cutting rules and the S-matrix does not satisfy unitarity
constraints.  We also
show that one loop amplitudes are unitary in the presence of only space
noncommutativity and the Feynman diagrams satisfy the cutting rules. 
We conclude in section $3$ with a discussion of our results 
 and their relation  to limits of 
string theory in electromagnetic field backgrounds.

\newsec{Unitarity of Noncommutative Scalar Field Theory} 

In this section we examine one loop diagrams of noncommutative
$\phi^3$ and $\phi^4$ 
theories to see if they satisfy constraints from unitarity. For
on-shell matrix elements unitarity 
implies that  
\eqn\unitary{
2 \,{\rm Im} \, M_{ab} = \sum_n M_{an} M_{nb} }
where $M_{ab}$ is the transition matrix element between states $a$ and
$b$. The sum over 
intermediate states on the right hand side includes phase space
integrations for each particle 
in $n$. Quantum field theories actually satisfy more restrictive
relations called generalized 
unitarity relations or cutting rules. These state that the imaginary
part of a Feynman diagram 
can be obtained by the following procedure: First, ``cut'' the diagram
by drawing a line through virtual lines such that the graph is severed
in two.  Next, wherever the cut intersects a virtual  
line, place that virtual particle on-shell by replacing the propagator
with a delta function:  
\eqn\os{
{1 \over p^2-m^2 + i \epsilon} \rightarrow -2\pi i \,\delta(p^2 -m^2) .} 
Summing over all cuts yields the imaginary part of the Feynman
diagram. Cutting rules are a generalization of \unitary\ to Feynman diagrams.
 Unitarity of the 
S-matrix \unitary\ follows from the cutting rules\foot{This assumes
of course that the poles of 
the propagators correspond to physical states. In gauge theories
unphysical states can 
propagate in loops and one must demonstrate that these states decouple
from the physical 
S-matrix. This will not be a concern for the scalar theories
considered in this paper.}.  Note that 
the cutting rules are more restrictive than the constraint of
unitarity since they apply to 
off-shell Green's functions as well as S-matrix elements.

We will first show that the two-point function of the noncommutative
$\phi^3$ theory does not 
obey the usual cutting rules when there is space-time noncommutativity
($\theta^{0i} \neq 0)$. In 
the case of space noncommutativity $(\theta^{0i} = 0,
\theta^{ij} \neq 0)$ the cutting 
rules are satisfied. Next, we consider $2\rightarrow 2$ scattering in
noncommutative $\phi^4$ 
theory.  The S-matrix is nonunitary at one-loop if $\theta^{0i} \neq
0$, but is unitary if the 
noncommutativity is only in the spatial directions.

It is somewhat surprising that Feynman diagrams of space-time
noncommutative theories do 
not obey the usual cutting rules. Since the Feynman rules for the
vertices of noncommutative 
theories are manifestly real functions of momenta, one would expect
that Feynman graphs 
could only develop a branch cut when internal lines go on-shell.  This
would imply that the 
imaginary parts of Feynman diagrams would be given by the same cutting
rules as ordinary 
commutative field theories. The resolution of this puzzle requires an
examination of the high 
energy behavior of the oscillatory factors that typically arise in
these theories. We will find that 
a necessary condition for one-loop Feynman integrals to converge in
these theories is that the 
following inner product 
\eqn\pop{ p \circ p = - p^\mu \theta^2_{\mu \nu} p^{\nu} , } 
be positive definite. This inner product is positive definite when
$p_\mu$ and $\theta_{\mu \nu}$ 
are analytically continued to Euclidean space. In Minkowski space $p
\circ p$ can be negative if 
$\theta^{0 i} \neq 0$. Feynman integrals can be defined via analytic
continuation, but the 
resulting amplitudes will develop branch cuts when $p \circ p < 0$.
These additional branch 
cuts are responsible for the failure of cutting rules and unitarity in
noncommutative theories 
with space-time noncommutativity.

For $p \circ p = 0$, the S-matrix does not suffer from lack of
unitarity, but is ill-defined because 
of infrared divergences. $p \circ p = 0$ is possible whether the
noncommutativity is space-time 
or space-space. Obviously an outstanding problem in noncommutative
field theory is to 
construct the infrared safe observables of the theory. This may
require all order resummation of 
infrared divergent terms in the perturbative series. We will not
atttempt to address this issue in 
this paper, and focus only on perturbative unitarity constraints for
matrix elements which do 
not suffer from infrared singularities.  

\eject

\subsec{Noncommutative $\phi^3$ two-point function} 

\ifig\cut{Generalized unitarity relation for $\phi^3$ two-point function.}
{\epsfxsize5in\epsfbox{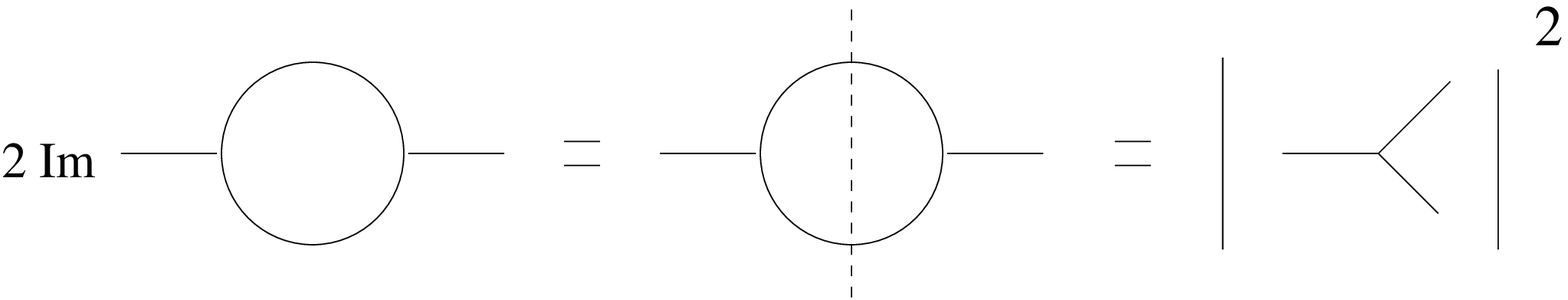}}

The cutting rule for the noncommutative $\phi^3$ theory two-point
function at lowest order is  
displayed in \cut. The propagators of fields in noncommutative field
theories are identical to those of commutative field theory.
The Feynman rule for the vertex in this theory is
\eqn\fr{ -i \,\lambda \, {\rm cos}\left( {k \wedge q \over 2}\right) ,
\,\,\,\,\,\, 
k \wedge q = k_\mu \theta^{\mu \nu} q_\nu . }
where $k$ and $q$ are any two of the momenta flowing into the
vertex. Because of 
conservation of momentum  and the antisymmetry of $\theta_{\mu \nu}$
it does not matter  
which two momenta are chosen.  The amplitude for the one loop diagram
appearing in \cut\ is: 
\eqn\lp{\eqalign{
i M ={\lambda^2 \over 2} \int {d^D l\over (2 \pi)^D} {1+{\rm cos}(p
\wedge l) \over 2} 
{1\over  l^2 -m^2 +i \epsilon }{1 \over (l+p)^2 -m^2 +i \epsilon}  \, ,}}
while the expression for the right hand side of \cut\ is 
\eqn\msq{
\sum |M|^2 ={\lambda^2\over 2}{1 \over (2 \pi)^{D-2}}\int {d^{D-1} k \over
2k_0}{d^{D-1} 
q \over 2q_0}  
\delta^D(p - k -q) {1+{\rm cos}(p \wedge k) \over 2}.}
The mass of the $\phi$ quanta is $m$ and $p$ denotes the external
momentum 
which is not required to be on-shell.
 In both \lp\ and \msq\ we
have used the identity 
${\rm cos}^2 x =(1+{\rm cos}(2x))/2$ to separate the planar and
nonplanar contributions
\nref\gonoka{A. Gonzalez-Arroyo and M. Okawa, ``The Twisted
Eguchi-Kawai Model:A Reduced Model For Large N Lattice Gauge Theory'',
Phys. Rev. {\bf D27} (1983)2397.}  
\nref\egunaka{T. Eguchi and R. Nakayama, ``Simplification of Quenching
Procedure For Large N Spin Models'', Phys. Lett. {\bf B122} (1983) 59.} 
\nref\filk{T. Filk, ``Divergences in a Field Theory on Quantum
Space'', Phys. Lett. {\bf B376} (1996) 53.}%
\gonoka-\filk. We 
will focus on the nonplanar terms since it is obvious that the planar
parts satisfy unitarity 
constraints.

First we compute the one loop graph. We combine denominators using
Feynman parameters 
then represent propagators via Schwinger parameters to obtain 
\eqn\feint{
M = {\lambda^2 \over 8} \int {d^D l_E\over (2 \pi)^D} \int_0^1 dx\hskip-3pt
\int_0^{\infty} \hskip-3pt d\alpha 
\, \alpha
\left( {\exp}(-\alpha(l_E^2+x(1-x) p_E^2 +m^2-i \epsilon) + i l_E
\wedge p_E) 
+ c.c. \right).}
We have performed the usual analytic continuation $l^0=i l_E^0, 
p^0=i p_E^0$. 
The subscript $E$ denotes Euclidean momenta.
In addition, if there is space-time noncommutativity we must
analytically continue $\theta^{0 i} 
\rightarrow -i \, \theta^{0 i}$. In the string theory realization this
can be easily undertood since $\theta^{0i}$ is related to a background
electric field. This continuation leaves the Moyal phase invariant.
Otherwise the phases appearing in \feint\ , $\exp(\pm i l_E\wedge
p_E)$, which render the integral finite, become $\exp(\pm l_E \wedge
p_E)$ and the integral 
is no longer convergent. Integrating over the loop momentum $l_E$ gives
\eqn\ans{
M={\lambda^2 \over 4} {1 \over (4 \pi)^{D/2}} \int_0^1 dx
\int_0^{\infty} d\alpha \, 
\alpha^{1-D/2}  \exp \left(-\alpha(x(1-x) p_E^2 +m^2-i \epsilon) -
{p_E \circ p_E \over  
4 \alpha} \right).}
We will now evaluate this integral for $D=3$ and $D=4$ space-time
dimensions and analytically continue back the answer to Minkowski space. The
amplitudes are given by
\eqn\thrdim{
M_{D=3}= {\lambda^2\over 32 \pi} \int_0^1 dx {\exp(- \sqrt{p \circ p(m^2 -p^2
x(1-x)- i \epsilon}) \over \sqrt{m^2-p^2 x(1-x) - i \epsilon})},}
and
\eqn\fourdim{
M_{D=4}= {\lambda^2 \over 32 \pi^2} \int_0^1 dx \,K_0( \sqrt{p \circ p(m^2
-p^2 x(1-x)- i\epsilon)}),} 
where $K_0$ is a modified Bessel function.
A crucial point to note is that the $\alpha$ integral is convergent
only if $p_E \circ p_E > 0$. 
For Euclidean momenta this is always true\foot{We will stay away from
the region where $p\circ p=0$ where infrared singularities appear.}. 
On the other hand, $p \circ
p$ need not be positive definite in Minkowski space when space-time is
noncommuting. Specifically, let us choose  
$\theta^{01}=-\theta^{10} = \Theta_E, \theta^{23}=-\theta^{32} =
\Theta_B$ with all other  
components vanishing. Then 
\eqn\popmk{ p\circ p = \Theta_E^2 (p_0^2 - p_1^2) + \Theta_B^2 (p_2^2
+ p_3^2).}
Therefore, in the case of only space noncommutativity $p \circ p$ is
positive definite but for space-time  
noncommutativity $p\circ p$ can be negative. This fact  has very
important consequences in the unitarity analysis.

We will now proceed to verify that the generalized unitarity relation
\unitary\ is satisfied for magnetic theories and violated for electric
field theories. First we compute 
the imaginary part of the Feynman diagram when $p^2>0$ and 
$p \circ p > 0$. It is then easy to show that 
\eqn\ImThree{\eqalign{
{\rm Im}\ M_{D=3}= {\lambda^2 \over 32 \pi}
\int_{(1-\gamma)/2}^{(1+\gamma)/2} dx \, 
{{\rm cos}(\sqrt{{p \circ p}} \sqrt{-m^2+p^2x(1-x)}) \over  
 \sqrt{-m^2+p^2x(1-x)}} \cr  
= {\lambda^2 \over 32\sqrt{p^2}} J_0\left({\gamma \sqrt{p^2 \, p\circ p} \over
2}\right) },} 
where $\gamma = \sqrt{1-4m^2/p^2}$.

\noindent
Using the fact that $\hbox{Im}\ K_0(-ix)={\pi\over 2}J_0(x)$, where
$J_0$ is a Bessel function, one obtains for $D=4$ space-time dimensions
\eqn\Imfour{ \eqalign{ 
{\rm Im}\ M_{D=4} = {\lambda^2 \over 64 \pi}
\int_{(1-\gamma)/2}^{(1+\gamma)/2} dx \, 
J_0(\sqrt{{p \circ p}} \sqrt{-m^2+p^2x(1-x)}) \cr  
= {\lambda^2 \over 32 \pi}  
{{\rm sin} ( \gamma\sqrt{p^2 \,p\circ p}/2) \over  \sqrt{p^2
\,{p \circ p}}} 
}.}

We will now evaluate the sum over final states \msq . The integrals
evaluate to
\eqn\msthree{
\sum |M_{D=3}|^2 = {\lambda^2\over 4}{1\over 8 \pi \sqrt{p^2} }\int_0^{2\pi}
d\theta \, {\rm cos}(p \wedge k) =  
{\lambda^2 \over 16 \sqrt{p^2}}   
J_0\left({\gamma \sqrt{p^2 \, p\circ p} \over 2}\right) }
and for $D=4$ space-time \msq\ gives
\eqn\msfour{
\sum |M_{D=4}|^2 = {\lambda^2 \over 
4}{\gamma \over 32 \pi^2}\int
d\Omega \, {\rm cos}(p \wedge k) = {\lambda^2 \over 16 \pi}  
{{\rm sin} (\gamma \sqrt{p^2 \,p\circ p}/2) \over  \sqrt{p^2
\,{p \circ p}}}.} 
We see that for $p\circ p > 0$ the generalized unitarity relation
\unitary\ is
satisfied. 

We will now consider the case $p \circ p < 0$. From \popmk\ it follows
that this configuration of
momenta can only exist in the presence of space-time
noncommutativity. Moreover $p^2$ must be negative so it corresponds to
space-like momentum. Then
\eqn\negpop{
{\rm Im}\ M_{D=3}
= {\lambda^2\over 32 \pi} \int_0^1 dx {{\rm sin}( \sqrt{|p \circ p|
(m^2 +|p^2| x(1-x)}) \over \sqrt{m^2+|p^2| x(1-x)})}}
and 
\eqn\negpopb{
{\rm Im}\ M_{D=4}={\lambda^2 \over 64 \pi} \int_0^1 dx \,J_0( \sqrt{|p
\circ p|(m^2 +|p^2| x(1-x))}).}
which are obviously nonzero. However,  the right hand side of the
equation in \cut\ is zero because 
energy-momentum conservation \msq\ forbids a particle with space-like
momenta to decay into two  
massive on-shell particles. Therefore, when $p \circ p < 0$, the
generalized unitarity relation \unitary\
is violated. 

Summarizing, we have shown that field theories with space-time
noncomutativity violate the equation in fig. 1 
and that
field theories with space noncommutativity satisfy
it for
arbitrary momenta.

\subsec{Noncommutative $\phi^4$ Scattering Amplitude}

Next we consider the $2 \rightarrow 2$ scattering amplitude in noncommutative 
$\phi^4$ theory. The Feynman rule for the 4-point vertex in this
theory is
\smallskip
$${ -i {\lambda \over 3} \bigg( \hskip-3pt{\rm cos}\left({p_1 \wedge
p_2\over 2}\right) \,\hskip-3pt {\rm 
cos}\left({p_3 \wedge p_4\over 2}\right) + \hskip-3pt 
{\rm cos}\left(\hskip-2pt{p_1 \wedge p_3\over 2}\hskip-2pt\right) \,\hskip-3pt {\rm
cos}\left(\hskip-2pt{p_2 
\wedge p_4\over 
2}\hskip-2pt\right) + \hskip-3pt{\rm cos}\left(\hskip-2pt{p_1 
\wedge p_4\over 2}\hskip-2pt\right) \,\hskip-3pt  
{\rm cos}\left(\hskip-2pt{p_2 \wedge p_3\over 2\hskip-2pt}\right)
\hskip-3pt\bigg)}$$ 
where the $p_i$ are momenta entering the vertex.

\ifig\onelp{One loop diagrams for $2 \rightarrow 2$ scattering}
{\epsfxsize4in\epsfbox{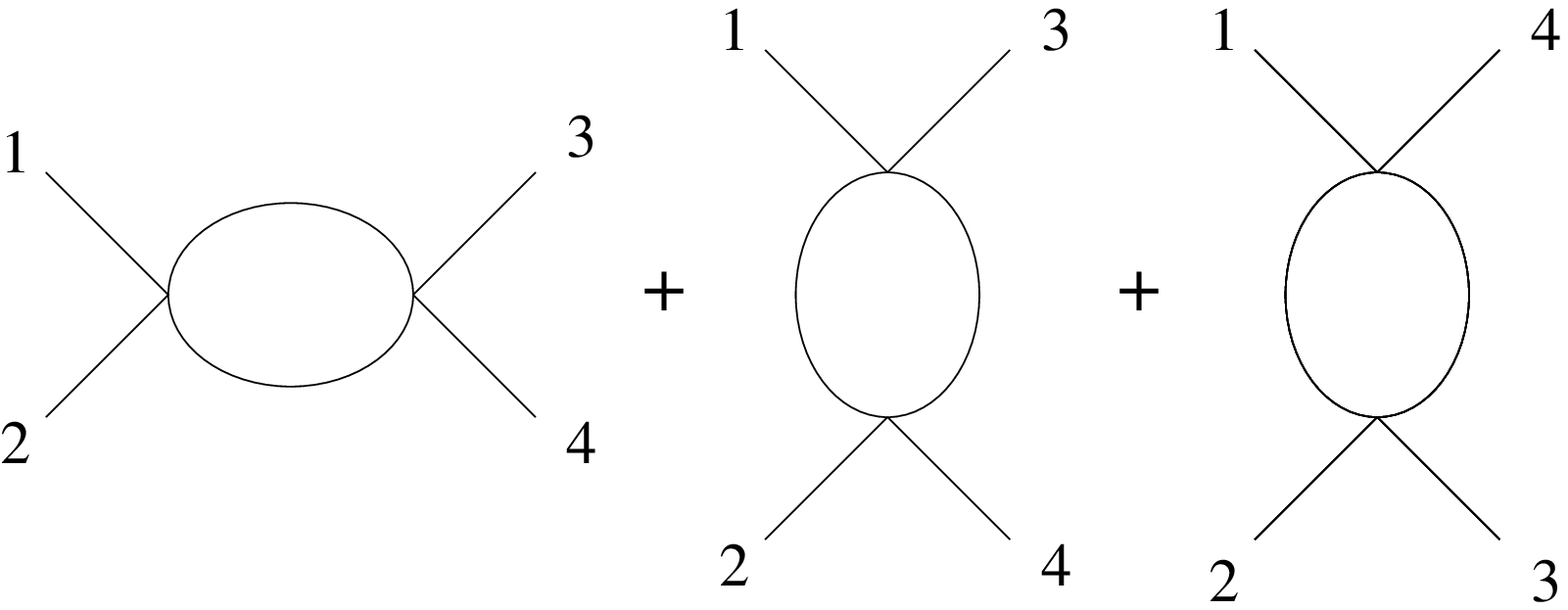}}

The one loop contribution to the $2 \rightarrow 2$ scattering
amplitude are shown in  
\onelp . Evaluating these graphs leads to rather complicated
expressions which involve integrals over modified Bessel functions,
but these  
simplify greatly if we expand the expressions in powers of
$\theta^{\mu\nu}$. The optical theorem \unitary\ for this S-matrix element has
to be true term by term in a power series in $\theta^{\mu\nu}$.
 The leading
contribution in $\theta$ to the right hand side of \unitary\ is 
the same as 
commutative $\phi^4$ theory, so 
\eqn\mspf{
\sum_n M_{p_1+p_2\rightarrow n}M^*_{n\rightarrow p_3+p_4} 
= { \gamma \lambda^2 \over 16 \pi} \Theta(s-4m^2)}
where $\Theta(x)$ is a step function, $\gamma = \sqrt{1-4m^2/s}$ and
$s = (p_1+p_2)^2 = (p_3+p_4)^2$. 

The leading contribution to the one loop 
 scattering amplitude is
\eqn\oneloop{\eqalign{
&M_{p_1+p_2 \rightarrow p_3+p_4} = \cr
&-{\lambda^2 \over 2 (4\pi)^2} \Bigg(
\int_0^1 dx\left[ {\rm ln}\left(1-{s \over m^2}x(1-x) \right) +  (s
\rightarrow t, s \rightarrow u ) 
\right] +{2\over 3} {\rm ln}\left( m^2 \over \mu^2 \right) + {\rm
const.} \cr 
&+{1\over 3} \left(\sum_{i=1}^4 {\rm ln}(m^2 \, p_i\circ p_i )
+{1\over 3} {\rm ln}(m^2p_{12}\circ p_{12})+{\rm ln}(m^2p_{13}\circ p_{13})
+{\rm ln}(m^2p_{14}\circ p_{14}) \right) \Bigg)}.}
Here we have defined $t =(p_1-p_3)^2, \, u=(p_1-p_4)^2, \, p_{12}=p_1+p_2,
\, p_{13}=p_1-p_3$, and 
$p_{14}=p_1-p_4$. The first line in \oneloop\ includes the contributions
present in ordinary 
$\phi^4$ theory, while the second includes logarithms that are unique
to the noncommutative 
theory. The first logarithm has an imaginary piece when $s>4 m^2$,
corresponding to threshold production of two $\phi$ particles, 
which gives the precise 
contribution so that \unitary\ is satisfied to leading order in
$\theta^{\mu\nu}$ with the right hand side of \unitary\ given by
formula \mspf . The noncommutative
logarithms ${\rm ln}(m^2 \, p_i 
\circ p_i)$ and ${\rm ln}(m^2 \, p_{12} \circ p_{12})$ do not
contribute an imaginary piece 
because $p_i, \,p_{12}$ are time-like, and hence $p_i \circ p_i,
\,p_{12} \circ p_{12}$ are 
positive definite. However, $p_{13}$ and $p_{14}$ are space-like and
therefore $p_{13} \circ 
p_{13}$ and $p_{14} \circ p_{14}$ can be negative if there is
space-time noncommutativity. In 
this case these logarithms  have imaginary parts which
violate the unitarity 
relation \unitary\ . Therefore, theories with space-time
noncommutativity do not have a unitary S-matrix. Moreover, since
$p\circ p>$ is always positive for space noncommutative theories there
are no new imaginary parts and the optical theorem is satisfied.

\newsec{Discussion}

In this paper we have investigated the unitarity of noncommutative
scalar field theories. The results we have obtained have a natural
interpretation in string theory. We have
shown that field theories with space noncommutativity appear to have
perturbatively unitarity S-matrix elements and satisfy the
generalized unitarity relations of field theory Green's functions. On
the other hand, theories with space-time
noncommutativity do not have a unitary S-matrix and do not satisfy the
cutting rules for Feynman diagrams.

We have done calculations for noncommutative scalar field
theories. Even though we have not checked the unitarity of
noncommutative 
gauge theories we have strong reasons to believe
that the same results still hold, that is the magnetic theories are
unitary while the electric theories are not. This is because the
structure of Feynman integrals is the same in gauge
and scalar theories. Both have oscillating phases in loop integrations.
After analytically continuing momenta and $\theta^{0i}$ to Euclidean
spacetime, then performing loop integrals one encounters integrals of the
form:
\foot{We are not including
integration over Feynman parameters which are not needed for the
argument.} 
\eqn\feyn{
A\sim\int{d\alpha}\ \alpha^{1-D/2}\exp^{-{p\circ p\over \alpha}},}
where $p$ denotes some external momentum. In the Euclidean theory
$p\circ p\geq 0$ so that  $1/p\circ p$ regulates \feyn\ and 
acts
like an ultraviolet cutoff which renders the integral finite. 
In the theory with only space
noncommutativity, the Minkowski expression for $p\circ p> 0$ is
never negative. The
only possible singularity arises when $\theta^{\mu\nu}p_\nu=0$ which
leads to characteristic  infrared singularities of noncommutative
field theories \mvs-\sunew. On the other hand when $\theta^{0i}\neq 0$, the
Minkowski expression for $p\circ p$ can be positive or negative so
that when one analytically continues the Euclidean answer, 
one finds branch cuts in the Feynman diagrams for Minkowski
$p\circ p < 0$. It is the presence of these extra branch cuts in the
loop diagrams of 
field theories with space-time noncommutativity that are responsible
for the failure  of the cutting rules  and lack of
a unitary S-matrix.

The fact that the magnetic gauge theories are unitary can be easily
understood from string theory. They provide the appropiate effective
description of a low energy limit of string theory in the presence of
a background magnetic field \cds\dh\sw . In this limit, 
all the massive open string
states and the closed strings decouple and the relevant degrees of
freedom for the description of the dynamics are the massless open
strings. One can build up the effective action for these modes from
string theory and
show that they are described by noncommutative field
theory. Therefore, we expect the field theory to be unitary since
string theory in this limit
can be appropriately described in terms of noncommutative field theory,
without the need of adding any further degrees of freedom. This is 
indeed what we have found from our field theory analysis.  

Field theories with space-time noncommutativity should appear from
studying string theory in the presence of a background electric
field. This follows from constructing the effective action of open strings in
this vacuum. One might expect, based on analogy with a background
magnetic field, that there is a similar limit of the string
dynamics which is described just by the electric field theory. However,
electric fields behave differently than magnetic fields in that they
lead to pair production of strings and these destabilize the vacuum if
the background electric field exceeds the upper critical value
$E_c$ \bur\bapo\nest. Consider for simplicity
 the electric field to be aligned
in the $x^1$ direction and the metric to be diagonal in the $x^0,x^1$
plane with each metric component given by $g$. Reality of the brane
action requires that the background electric field on the brane
satisfy
\eqn\back{
E\leq E_c, \qquad \hbox{where}\quad E_c={g\over 2\pi\alpha^\prime}.}
The open strings see a diagonal metric along the $x^0,x^1$
plane given by $G$ and a noncommutative parameter
$\theta^{0i}=\theta$. In terms of the metric $g$ and the background
electric field, these parameters are related by the following formula
\sstb\gmms\br
\eqn\expresa{
\alpha^{\prime}G^{-1}={1\over 2\pi}{E\over E_c}{\theta}.}
In order to obtain a field theory of only the massless modes one has
to go to the point particle limit $\alpha^\prime\rightarrow 0$. From
formula \expresa\ this implies that, for finite $G$, that the
noncommativity parameter must vanish. Therefore, if one wants a
truncation of the full string theory to the theory of only the
massless open string modes this can be done but the description of
these modes is given by conventional field theory and not
noncommutative field theory. Thus, we expect the conventional field
theory description to be unitary and it is. Clearly, in order to have
a finite noncommutativity parameter $\theta$, $\alpha^\prime$ must be
kept finite. This is a string theory and not a field theory. Moreover,
since $\theta\sim \alpha^\prime$ there is no scattering process in
this string theory which is accurately described only by
noncommutative field theory. For scattering processes involving
massless open strings
of characteristic energy
$E\ll \theta^{-1/2}$ conventional field theory is a proper
description.  Noncommutative field theory becomes relevant for energies
of the order of  $E\sim\theta^{-1/2}$. However, since 
$\theta\sim\alpha^\prime$, the energy scale where
 noncommutativity becomes relevant is
precisely the energy scale at which the massive open string states
cannot be neglected. Thus, there is no regime in which space-time
noncommutative field theory is an appropiate description of string
theory. Whenever space-time noncommutative field theory 
 becomes relevant massive open string states
cannot be neglected. This gives a very strong indication that
noncommutative field theories with space-time noncommutativity are not
unitary, since they do not have all the relevant degrees of
freedom necessary for an approximate description of a unitary string
theory. This is what we found from our field theoretic analysis.

 Recently, it has been noticed by several groups \sstb\gmms\
that it is possible to
define  a limit of string theory in a background electric field in
which the full tower of open string states decouple from the closed
strings \sstb\gmms. In this limit the background electric field is
sent to its critical value (see
\nref\gkp{S. Gukov, I.R. Klebanov and A.M. Polyakov, `` Dynamics of (n,
1) Strings'', Phys. Lett. {\bf B423} (1998) 64, hep-th/9711112.}%
\gkp\ for previous analysis of this limit). There is a very simple
way of showing that indeed closed strings decouple in this
limit. Quantization of  open strings with
the modified boundary conditions due to the electric field 
lead to familiar looking mode
expansions \clny\bapo\ for the light-cone coordinates
$X^{\pm}$. In the limit that $E\rightarrow E_c$, the waves on the string for
the $X^{\pm}$ directions become chiral, that is, they are either purely
right moving or left moving waves. Therefore, in this limit, 
it is impossible for
such an open string to become  a closed string, since closed strings
require waves which are left moving and right moving.

There is still a lot to learn about noncommutative theories, both
field theories with space noncommutativity as well as the recently
discovered decoupled open string theories with space time 
noncommutativity \sstb\gmms. The infrared divergences in the
magnetic theories for $p\circ p=0$ certainly need to better understood
within a field theory framework. These theories, as they stand, have
no infrared safe observables and the S-matrix is ill-defined. Perhaps
nonperturbative input will be required to address this problem. 

\medskip
\medskip
\bigskip
\centerline{\bf Acknowledgments}

We would like to thank Steve Giddings, Djordje Minic, Hirosi Ooguri, 
Mark Wise and 
Edward Witten for helpful discussions and Hirosi Ooguri for carefully
reading of the manuscript. J.G. and T.M.
are supported in part by the DOE under grant no. DE-FG03-92-ER
40701.

\listrefs

\end